\PassOptionsToPackage{pdftex}{hyperref}
\documentclass[fleqn]{2023SCGE}
\usepackage{graphicx}
\usepackage{booktabs}
\usepackage{amsmath,amsfonts,amssymb,bm,upgreek,mathrsfs}
\usepackage{multirow}
\usepackage{xspace}
\usepackage{xcolor}
\usepackage{etoolbox}
\usepackage{float}


\newcommand{\rpar}{r_{\parallel}}
\newcommand{\rperp}{r_{\perp}}
\newcommand{\blya}{b_{\rm Ly\alpha}}

\newcommand{\hmpc} {h^{-1}\mathrm{Mpc}}

\newcommand{\PD} {P_{\rm 1D}}

\begin{document}

\ensubject{subject}

\ArticleType{Article}
\SpecialTopic{}
\Year{2025}
\Month{November}
\Vol{66}
\No{1}
\DOI{??}
\ArtNo{000000}
\ReceiveDate{}
\AcceptDate{}

\title{Low-redshift $3$D Lyman-$\alpha$ Forest Correlations with China Space Station Telescope}{Low-redshift $3$D Lyman-$\alpha$ Forest Correlations with China Space Station Telescope}

\author[1]{Ting Tan}{{ting.tan@cea.fr}}
\author[2]{Huanyuan Shan}{}
\author[1]{Eric Armengaud}{}

\AuthorMark{Tan T}
\AuthorCitation{Tan T, Shan H Y, Armengaud E}

\address[1]{CEA, IRFU, Universit\'e Paris-Saclay, F-91191 Gif-sur-Yvette, France}
\address[2]{Shanghai Astronomical Observatory, Chinese Academy of Sciences, Shanghai, China}

\abstract{While the Lyman-$\alpha$ (Ly$\alpha$) forest traces the large-scale matter distribution over a wide range of redshift, its three-dimensional (3D) clustering at $z<2$ has not yet been measured. We investigate the prospects for measuring low-redshift Ly$\alpha$ correlations with the UV slitless spectroscopic instrument of the China Space Station Telescope (CSST). We construct mock CSST quasar spectra that reproduce the expected survey depth, spectral resolution and noise properties, and derive Ly$\alpha$ auto-correlation functions and cross-correlations with quasars (QSO) and emission-line galaxies (ELG) in the range $1.1<z<2.0$. We then interpret these three-dimensional correlation functions with a standard anisotropic redshift-space clustering model and obtain forecast constraints on the Ly$\alpha$ and tracer parameters. At an effective redshift $z_{\rm eff}=1.59$ (1.58 for ELGs), the Ly$\alpha$ bias parameters will be measured with a 10-30\% precision, depending on priors on other tracer’s
biases. We also forecast a marginal $2.5\sigma$ ($3.7\sigma$) detection of the BAO feature, corresponding to a $\sim10\%$ (7\%) constraint on the isotropic BAO scale, from the combination of Ly$\alpha$ auto- and Ly$\alpha$–QSO (ELG) cross-correlations. These results show that CSST can provide the first three-dimensional characterization of the low-redshift Ly$\alpha$ forest and a complementary Ly$\alpha$-based BAO measurement at $z<2$, helping to link galaxy clustering surveys with high-redshift Ly$\alpha$ forest studies.}

\keywords{Baryon Acoustic Oscillations, Lyman-$\alpha$ forest, Cosmology}
\PACS{98.80.-k, 98.80.Es}

\maketitle

\begin{multicols}{2}
\section{Introduction}
The Lyman-$\alpha$ (Ly$\alpha$) forest, defined as the ensemble of absorption features from neutral hydrogen (H\,\textsc{i}) in quasar spectra arising in the diffuse intergalactic medium (IGM), traces mildly non-linear matter fluctuations over a wide range of redshift and spatial scales \cite{rauch1998lya,weinberg2003lya}. In the fluctuating Gunn--Peterson approximation, the transmitted flux is tightly correlated with the underlying density field \cite{mcdonald2006lyalpha}, so Ly$\alpha$ forest statistics provide constraints on the matter power spectrum, redshift-space distortions,
\Authorfootnote
and the thermal and ionization state of the IGM.

At redshifts $z> 2$, the Ly$\alpha$ forest has been mapped extensively by large optical spectroscopic surveys, including the Baryon Oscillation Spectroscopic Survey (BOSS) \cite{dawson2013boss}, the extended Baryon Oscillation Spectroscopic Survey (eBOSS) \cite{dawson2016eboss}, and the Dark Energy Spectroscopic Instrument (DESI) \cite{desi2016instrument,desi2016science}. These programs have provided precise measurements of both the one-dimensional Ly$\alpha$ flux power spectrum \cite{palanque2013p1d,chabanier2019p1d}, and the baryon acoustic oscillations (BAO) from the 3D Ly$\alpha$ auto-correlation and its cross-correlation with quasars \cite{slosar2013lya,delubac2015baryon,des2020completed,desi2023lya,desi2024lya,abdul2025desi}. 


By contrast, UV spectroscopy is required to observe the low-redshift Ly$\alpha$ forest, which remains much less well characterized than at $z>2$. 
Direct constraints on the Ly$\alpha$ one-dimensional flux power spectrum, $P_{\rm 1D}$, at $z<2$ are still scarce and are based on comparatively limited samples of space-based UV spectra; as one example, the $\PD$ measurement from HST/COS data at $z<0.5$\cite{khaire2019p1d}. 
No three-dimensional (3D) measurement of Ly$\alpha$ forest clustering, bias or redshift-space distortions has yet been obtained in this low redshift range, mainly because of the absence of a wide-area, UV-capable spectroscopic survey with sufficiently high quasar surface density.

The China Space Station Telescope (CSST) is designed to conduct a wide-field slitless spectroscopic survey with three low-resolution grism bands, including the GU channel covering 255--410\,nm \cite{gong2019cosmology,gong2025introduction}. This band records Ly$\alpha$ absorption from quasars in the range $1.1< z< 2.4$, and the planned wide survey will cover $\sim 1.7\times10^4\,\mathrm{deg}^2$ with a high surface density of quasars and emission-line galaxies (ELGs) \cite{wen2024csst}. Although the spectral resolution ($R\simeq 200$--250) and per-pixel signal-to-noise ratio are modest, the combination of large area and abundant sightlines makes CSST a promising facility for low-redshift Ly$\alpha$ forest cosmology.

In this work we present end-to-end forecasts for 3D Ly$\alpha$ forest measurements with CSST in the redshift interval $1.1<z<2.0$. By constructing CSST quasar catalogs and Ly$\alpha$ forest mocks, as well as a Euclid-like \cite{laureijs2011euclid,blanchard2020euclid} emission-line galaxy (ELG) sample, we compute Ly$\alpha$ forest auto-correlations and cross-correlations with quasars and ELG positions.
We interpret these measurements with a standard anisotropic redshift-space clustering model and derive expected constraints on the Ly$\alpha$ and tracer parameters.
At $z_{\rm eff}=1.59$ (1.58 for ELGs), the Ly$\alpha$ bias parameters can be constrained at the 10 to 30\% level, depending on priors on the tracer biases, and the combination of auto- and cross-correlations yields a $2.5\sigma$ ($3.7\sigma$) BAO detection with quasars (ELGs), corresponding to a $\sim10\%$ (7\%) constraint on the isotropic BAO scale.
These measurements will provide the first 3D Ly$\alpha$ forest clustering at low redshift, helping to link galaxy clustering surveys and high-redshift Ly$\alpha$ studies, thereby strengthening constraints on the IGM properties and expansion history during the critical transition from matter-dominated to dark-energy-dominated epochs.

The structure of this paper is as follows.
Sec.~\ref{sec:mock_generation} describes the CSST survey configuration and the construction of Ly$\alpha$ forest mock data.
Sec.~\ref{sec:measure_cf+fitting} presents the measurement of the 3D correlation functions and the model fitting methodology.
Sec.~\ref{sec:results} summarizes the forecast constraints on Ly$\alpha$ and BAO parameters, and Sec.~\ref{sec:discussion_conclusion} discusses the implications for low-redshift Ly$\alpha$ forest cosmology and future extensions.

\section{Simulated CSST Ly$\alpha$ sample}
\label{sec:mock_generation}
\subsection{Survey setting}

The Chinese Space Station Telescope (CSST) will conduct a large-area slitless spectroscopic survey in three bands: GU (255--410\,nm), GV (400--640\,nm), and GI (620--1000\,nm), with a resolving power of $R\simeq200$--250 in each band \cite{gong2019cosmology,gong2025introduction}. In this study we only consider observations in the GU band. In its default wide-field mode, the survey will cover 16,759 deg$^2$ (full sky with realistic galactic and ecliptic masks) over ten years, reaching a source depth of $m_{\mathrm{AB}}=23.2$ with $ 4\times150$\,s exposures per field. A complete description of the survey design and performance can be found in the CSST science white paper \cite{gong2025introduction}.


\subsection{Mock QSO and ELG catalogs}
\label{sec:Generation_QSO_ELG_mock}

\begin{figure}[H]
    \centering
    \includegraphics[width=\linewidth]{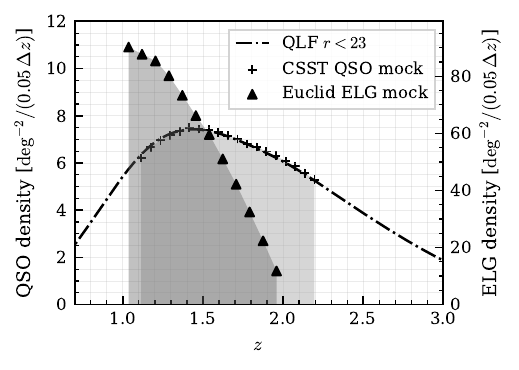}
    \caption{Redshift distribution of quasars ($1.1<z_{\rm{QSO}}<2.2$) and ELGs ($1.0<z_{\rm{ELG}}<1.9$) used in our mock, compared to QSO luminosity function (QLF) predictions for  magnitude limit $r < 23$.}\label{fig:qso_dndz}
\end{figure}

To generate correlated large-scale structure fields and tracer catalogues we use the publicly available \textsc{CoLoRe} code \cite{ramirez2022colore}. \textsc{CoLoRe}\footnote{\url{https://github.com/damonge/CoLoRe}} produces fast lognormal realizations of the matter density field over large cosmological volumes and populates biased tracers on the light cone according to specified redshift distributions and bias prescriptions, yielding self-consistent mock catalogs and density fields. This framework has been extensively used to construct mock data sets for eBOSS and DESI Ly$\alpha$ forest analyses and related clustering studies \cite{etourneau2024mock,herrera2025synthetic}.

In our implementation, \textsc{CoLoRe} generates a full-sky light cone from a periodic box of comoving side length $L_{\rm box}=8092\,h^{-1}\mathrm{Mpc}$ sampled on a $1024^3$ grid (real-space resolution $d x \simeq 7.9\,h^{-1}\mathrm{Mpc}$), covering the redshift range $0<z<2.4$. In the following analysis we focus on Ly$\alpha$ forests from quasars within the redshift interval $1.1 < z < 2.0$ (observed-frame wavelengths $\lambda \in [2550,3650]$\,\AA, fully covered by the GU band) and tracer quasars with $1.1 < z < 2.2$ used for cross correlations. We focus on this redshift coverage since it is complementary to the higher-redshift regime ($z>2$) typically targeted in eBOSS and DESI Ly$\alpha$ forest analyses.

We draw quasar and ELG catalogs from the same underlying matter realization so that their density fields are correlated. For quasars, the CSST-like catalog is generated using the QSO luminosity function (QLF) of \cite{palanque2016extended} and a redshift-dependent linear bias calibrated on eBOSS clustering measurements~\cite{laurent2017eboss}:
\begin{equation}
    b_{\rm QSO}(z) = 0.568 + 0.278\,(1+z)^2.
    \label{eq:bqso}
\end{equation}
Based on the spectroscopic depth of the CSST wide survey ($r<23.2$), we select quasars with $r<23.0$ in the targeted redshift range, therefore roughly taking into account the expected $\sim90\%$ completeness in quasar identification \cite{pang2025pilot}. The resulting mock redshift distribution of quasars (black cross points in Figure~\ref{fig:qso_dndz}) closely follows QLF predictions for $r<23$, with a surface density of $159\,\mathrm{deg}^{-2}$.

In addition, Euclid, DESI and CSST will all provide large ELG samples in the overlapping redshift. To measure the corresponding Ly$\alpha$–ELG cross-correlations, we generate an ELG catalog simultaneously with the quasars in \textsc{CoLoRe}, assuming a Euclid-like spectroscopic galaxy density, redshift distribution (shown in Figure~\ref{fig:qso_dndz}) and clustering bias \cite{pozzetti2016ha,blanchard2020euclid}:
\begin{equation}
    b_{\rm ELG}(z) = 1.46 + 0.68\,(z-1).
    \label{eq:belg}
\end{equation}

\subsection{Simulated Ly$\alpha$ transmission field}
We generate Ly$\alpha$ forest transmission fields with \textsc{LyaCoLoRe}
\cite{farr2020lyacolore}, which post-processes the Gaussian density and
velocity skewers from \textsc{CoLoRe} to produce transmitted-flux
skewers on the light cone. Following \cite{farr2020lyacolore},
\textsc{LyaCoLoRe} first restores missing small-scale line-of-sight
fluctuations by adding an independent Gaussian component
$\delta_\epsilon$ with a calibrated one-dimensional power spectrum, and
sets its amplitude through a redshift-dependent parameter
$\sigma_\epsilon(z)$. The resulting Gaussian field $\delta_G$ is then
mapped to a lognormal field $\delta$ and converted into a
real-space Ly$\alpha$ optical depth using a fluctuating
Gunn--Peterson approximation,
\begin{equation}
\tau(z)\;=\;\tau_0(z)\,\bigl[1+\delta(z)\bigr]^{\alpha(z)},
\end{equation}
where $\tau_0(z)$ controls the mean absorption and $\alpha(z)$ the
density--optical-depth response. Redshift-space distortions are then
included using the line-of-sight velocity field provided by
\textsc{CoLoRe}, yielding a redshift-space optical depth and
the transmitted flux
$F\;=\;\exp\!\left[-\tau\right]$. The triplet $\{\alpha,\tau_0,\sigma_\epsilon\}$ therefore controls, respectively, the response of the optical depth to density fluctuations, the mean effective optical depth, and the small-scale fluctuations of the Ly$\alpha$ forest. These functions are tuned \cite{farr2020lyacolore} so that the resulting mock spectra reproduce the measured Ly$\alpha$ $\PD$, flux probability distribution function and large-scale clustering at $z> 2$.

In this work, we do not attempt to calibrate again these parameters: instead, we keep the same parameters as in~\cite{farr2020lyacolore}, extrapolating the functions $\alpha(z)$, $\tau_0(z)$ and $\sigma_{\epsilon}$ down to $z=1$, in order to model the Ly$\alpha$ absorption accessible to the GU band. We then generate Ly$\alpha$ transmission fields over the observed-frame wavelength range $\lambda \in [2550,3650]$$\text{\AA}$ with $\Delta\lambda=2\text{\AA}$. Note that for the purpose of this forecast, only Ly$\alpha$ transmission fields are produced, while other astrophysical contaminants, in particular high column density systems (HCDs) and metal lines are not taken into account.

In the following, we check that the extrapolation of the \textsc{LyaCoLoRe} mock tuning for redshifts $z<2$ is physically sound. For that purpose, we compare their mean transmitted flux, large-scale Ly$\alpha$ bias and one-dimensional power spectrum $\PD$ with some existing measurements or simulations.

\begin{figure}[H]
\centering
\includegraphics[width=\linewidth]{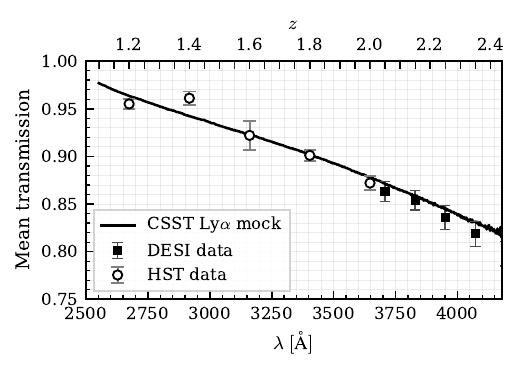}
\caption{Mean transmitted flux $\langle F(z) \rangle$ as a function of redshift in our \textsc{LyaCoLoRe} mocks (solid curve), obtained by averaging the Ly$\alpha$ transmission in narrow redshift slices. The hollow symbols show HST measurements (without strong absorption systems) from \cite{kirkman2007continuous}, and the solid points indicate the latest measurements from DESI at high redshift \cite{turner2024new}.}
\label{fig:mean-flux}
\end{figure}

We first compute the mean transmitted flux $\langle F(\lambda)\rangle$ from the mock transmissions. Figure~\ref{fig:mean-flux} shows the resulting mock mean transmitted flux, together with the HST measurements of~\cite{kirkman2007continuous}, as well as an example measurement at higher redshift from DESI~\cite{turner2024new}. Our mock transmission is consistent with the existing low-redshift HST data over the range $1<z<2$ relevant for CSST.

\begin{figure}[H]
    \centering
    \includegraphics[width=\linewidth]{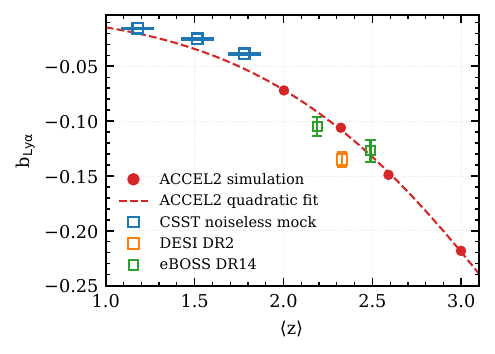}
    \includegraphics[width=\linewidth]{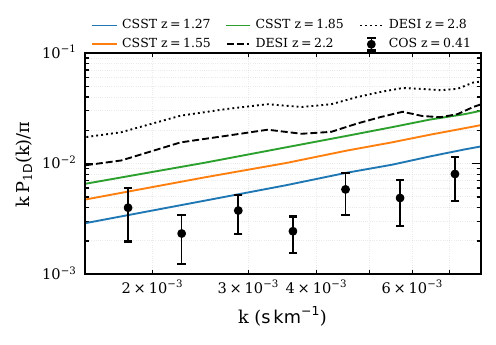}
    \caption{
    \emph{Top:} Redshift evolution of the Ly$\alpha$ forest bias. Red circles show the ACCEL2 hydrodynamical simulation results \cite{chabanier2024accel2}, more precisely from the fit to the \texttt{ACCL2\_L160R25} box. The red dashed curve is a quadratic fit. Blue squares with error bars are the Ly$\alpha$ biases measured from our simulated CSST transmission field in three redshift bins. Orange and green squares show the measurements from DESI DR2 ($z_{eff}=2.33$) and eBOSS DR14 ($z_{eff}=2.19$ and $2.49$, table B1 in \cite{de2019baryon} assuming no correlations between bias parameters) respectively.
    \emph{Bottom:} One-dimensional flux power spectrum $P_{\rm 1D}(k)$ in the same redshift bins, compared with measurements from DESI and HST/COS.}
    \label{fig:lya_bias_csst_accel2}
\end{figure}
We next assess the realism of the large-scale Ly$\alpha$ bias in the same redshift range. We divide the Ly$\alpha$ mock sample into three bins,
$1.0 < z < 1.5$, $1.5 < z < 2.0$ and $1.9 < z < 2.25$, then 
we measure the Ly$\alpha$ auto-correlation function and fit it to a simple bias model in each bin. These steps are done using the same pipeline as presented in detail later in Section~\ref{sec:measure_cf+fitting}. The fit is carried out varying only the Ly$\alpha$ bias and redshift-space distortion parameters: the resulting best-fit values of $\blya(z)$ are shown as blue squares in the upper panel of Fig.~\ref{fig:lya_bias_csst_accel2}. To our knowledge there is no $\blya$ measurement at $z<2$. For comparison we show bias measurements at higher redshifts from DESI \cite{abdul2025desi} and eBOSS \cite{de2019baryon}. In addition, we show bias calculations from the ACCEL2 hydrodynamical simulation \cite{chabanier2024accel2}, a state-of-the-art cosmological hydrodynamical simulation of the intergalactic medium tailored to Ly$\alpha$ forest studies. A quadratic fit, extrapolated to low redshift (red dashed curve) indicates that the low-redshift bias of our simulated transmission field is physically sound, and in agreement with this extrapolation. Correlated to the evolution of the mean transmission, the Ly$\alpha$ bias at $z<2$ is much lower than for $z>2$. In the absence of measurement, it is impossible to properly control the validity of our extrapolation, but it is conservative in the sense that we slightly under-estimate $\blya$ with respect to the ACCEL2-based extrapolation.

Finally, we compute $\PD$ of our simulated sample using the FFT estimator \cite{ravoux2025desi}, 
\begin{equation}
    P_{\rm 1D}(k_\parallel,z)
    = \left\langle \big|\tilde{\delta}_F(k_\parallel,z)\big|^2 \right\rangle,
\end{equation}
where $\delta_F$ is the fluctuation of transmission fields that will be defined later in Eq.~\ref{equa:delta}. The lower panel of Fig.~\ref{fig:lya_bias_csst_accel2} shows $P_{\rm 1D}$ estimated from the three CSST redshift bins, compared with the $\PD$ measurements of DESI \cite{karaccayli2025desi} and with one of the low-redshift COS measurements \cite{khaire2019p1d}. The overall shape and amplitude indicate that the $\PD$ of our simulated transmission field is physically meaningful.

Taken together, these tests indicate that the \textsc{LyaCoLoRe} transmission fields, using parameters simply extrapolated from high-redshift tuning, provide a plausible description of the low-redshift Ly$\alpha$ forest for the purposes of our CSST forecast.

\subsection{Instrumental resolution and noise}
\label{sec:resolution-noise}
\begin{figure}[H]
\centering
\includegraphics[width=\linewidth]{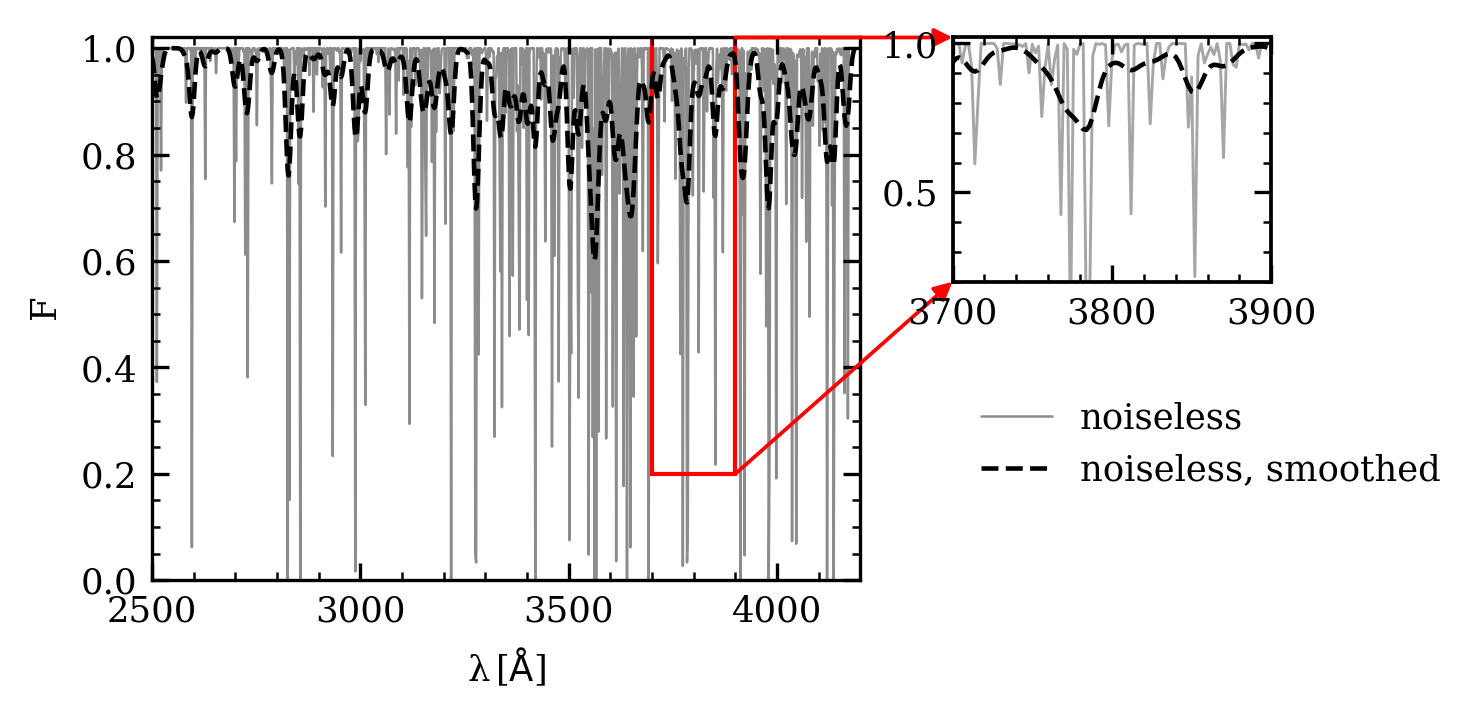}
\caption{Example simulated transmission skewer, with (black, dashed) and without resolution smoothing (gray, solid).
}
\label{fig:forest_spectrum}
\end{figure}
To emulate the spectral resolution of CSST spectroscopic observations, we post-processed the simulated Lyman-$\alpha$ forest transmission fields generated by \texttt{LyaCoLoRe}. For each line of sight, the high-resolution transmitted flux $F(\lambda)$ is convolved with a Gaussian kernel to simulate the line-spread function of an instrument with a resolving power $R= 241$ \cite{wen2024csst}. 

The observed transmission flux $F_{\mathrm{obs}}(\lambda)$ is then given by the convolution:

\begin{equation}
F_{\mathrm{obs}}(\lambda) = \int d\lambda' \, F(\lambda') \, G(\lambda - \lambda'; \sigma_\lambda),
\end{equation}
where $G(\lambda - \lambda'; \sigma_\lambda)$ is a normalized Gaussian function with standard deviation $\sigma_\lambda = \lambda / (2.355\, R)$, 
Figure~\ref{fig:forest_spectrum} (left plot) shows an example transmission in our mocks before and after spectral smoothing.

After spectral smoothing, we further incorporated observational noise at the level of the transmitted flux fluctuation field $\delta_F(\lambda)$, defined as:

\begin{equation}
\delta_F(\lambda) = \frac{F(\lambda)}{\langle F(\lambda) \rangle} - 1,
\label{equa:delta}\end{equation}
where $\langle F \rangle$ is the mean flux transmission. At this stage, delta fields $\delta_F(\lambda)$ are rebinned to $\Delta\lambda=8$\,\AA. To simulate pixel-level noise consistent with the expected observational data quality, we perturbed the $\delta$ values with additive Gaussian noise, which is sampled from $\mathcal{N}(0, \sigma_{\delta})$:
\begin{equation}
\delta_{\mathrm{noisy}}(\lambda) = \delta_F(\lambda) + \mathcal{N}(0, \sigma_{\delta}), 
\end{equation}
with
\begin{equation}
\sigma_{\delta} = \frac{\sigma_F}{\langle F \rangle}\approx\frac{1}{\mathrm{SNR}_{F}}\approx \frac{1}{\mathrm{SNR}_{f}}.\label{equa:sigma_delta}
\end{equation}

\begin{figure}[H]
\centering
\includegraphics[width=\linewidth]{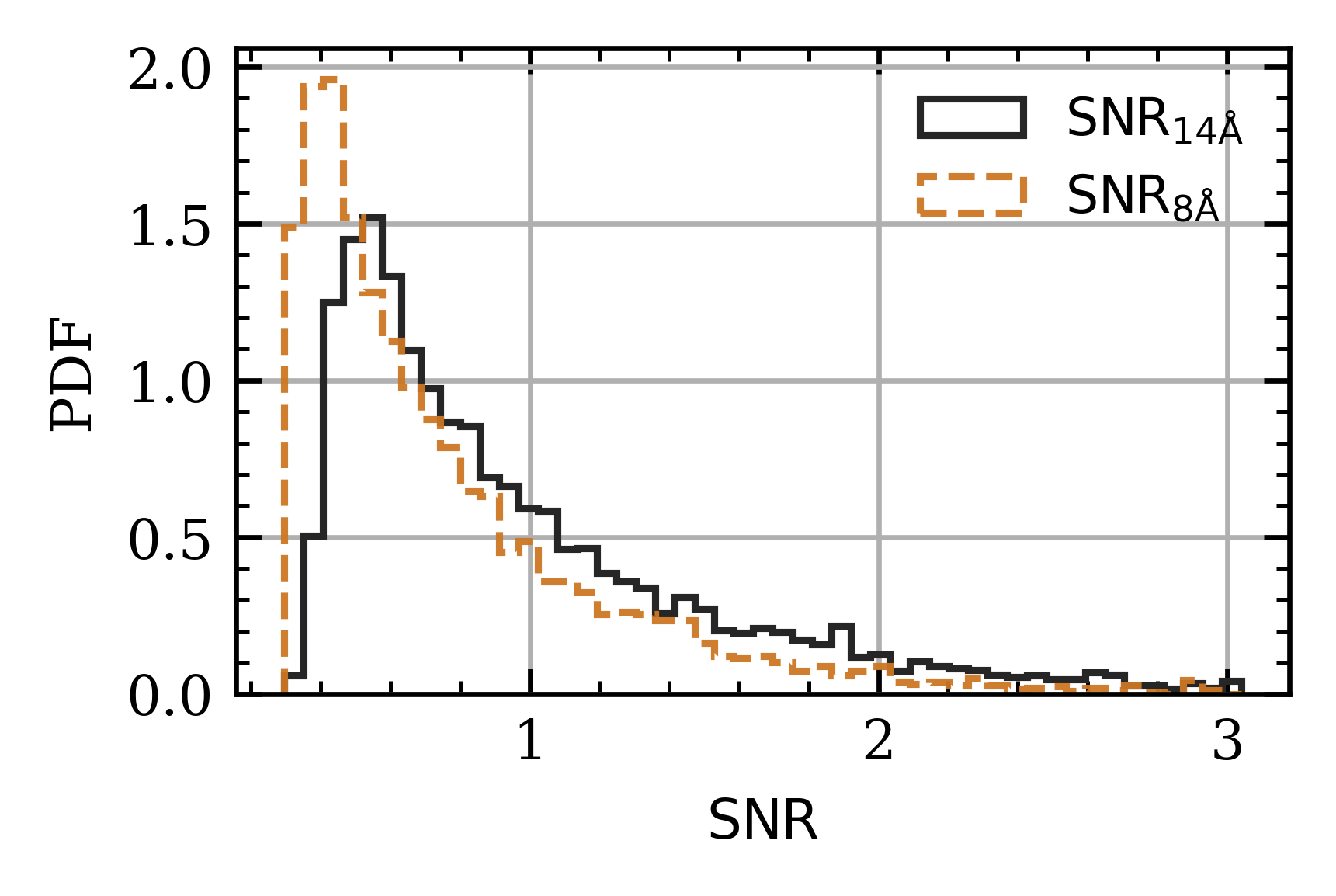}
\caption{Distribution of the empirically estimated SNR values $\mathrm{SNR}_{8\text{\AA}}$ (solid line) and $\mathrm{SNR}_{14\text{\AA}}$ (dashed line) for our mock Lyman-$\alpha$ sample.}
\label{fig:snr_hist}
\end{figure}

\noindent In this expression, $\mathrm{SNR}_{F}$ is the signal-to-noise ratio of the measured transmission flux. The latter is related to the measured quasar flux $f(\lambda) = C(\lambda)\,F(\lambda)$ where $C(\lambda)$ is the unabsorbed quasar continuum. We assume here that $C$ can be extracted perfectly, so that $\mathrm{SNR}_{F}$ is equal to the flux signal-to-noise $\mathrm{SNR}_f$. From simulated CSST galaxy spectra in the GU band presented in~\cite{wen2024csst}, we derived the following simple prescription:
\begin{equation}
\log_{10}[\mathrm{SNR}_{14\text{\AA}}(f)]
= -0.261+0.469\log_{10}\left(\frac{f}{f_0}\right),
\label{eq:snr_flux_fit}
\end{equation}
where $f$ is the object's average flux in the GU band, and $f_0 = 10^{-17}\,\mathrm{erg\,s^{-1}\,cm^{-2}}$. This prescription is for spectra with average bin width $\Delta\lambda=14\,\text{\AA}$ used in~\cite{wen2024csst} to compute SNR, and we apply it to an 8~\AA~bin width assuming the scaling $\mathrm{SNR}\propto \sqrt{\Delta\lambda}$. For each simulated line-of-sight, we compute $f$ using the quasar spectral modelling of the \texttt{simqso} package\cite{mcgreer2021simqso}, together with the QLF-based magnitude distribution as presented in Sec.~\ref{sec:Generation_QSO_ELG_mock}. The distribution of derived SNR is shown in Fig.~\ref{fig:snr_hist}: for our binning $\Delta\lambda=8\mathrm{\AA}$, its median value is estimated as 0.6.

\section{Correlation Function and BAO Fitting}\label{sec:measure_cf+fitting}

\subsection{Measurement of correlation functions}
\label{sec:measure_cf}

To compute the two-point statistics of the Ly$\alpha$ forest, we work directly with the Ly$\alpha$ transmission fluctuation fields $\delta(\lambda)$ defined in Eq.~\ref{equa:delta}, and do not perform any quasar continuum fitting. The correlation functions are then estimated from pairs of transmission fluctuations $(\delta_q(\lambda),\delta_{q'}(\lambda'))$, which are binned according to their angular and redshift separations $(\Delta\theta,\Delta z)$, where $q$ and $q'$ label different quasar sightlines.
We use a fiducial cosmological model \cite{aghanim2020planck} to transform these angular and redshift separations
$(\Delta z,\Delta\theta)$ into co-moving separation distances $(\rperp,\rpar)$:
\begin{equation}
\begin{aligned}
r_{\parallel} &= \left[ D_c(z_i) - D_c(z_j) \right] \cos\left( \frac{\Delta \theta}{2} \right), \\
r_{\perp} &= \left[ D_M(z_i) + D_M(z_j) \right] \sin\left( \frac{\Delta \theta}{2} \right).
\end{aligned}
\end{equation}
Here $D_c(z) = \int_0^z \frac{dz}{H(z)}$ is the line-of-sight comoving distance, while $D_M(z)$ denotes the transverse comoving distance; since our fiducial cosmology is spatially flat, $D_M(z)=D_c(z)$.

Using the \texttt{PICCA}\footnote{\url{https://github.com/igmhub/picca}} software \cite{du2021picca}, we compute the forest auto-correlation function
and quasar-forest cross-correlation 
in bins, labelled $A$, with transverse and longitudinal sizes of $4\hmpc$:
\begin{equation}
	\xi_{A}^{auto} = \frac{
    \sum\limits_{(i,j) \in A} w_{i}w_{j} \, \delta_{i}\delta_{j}
    }{
	\sum\limits_{(i,j) \in A} w_{i}w_{j}
    }
    \hspace*{3mm}
    \xi_{A}^{cross} = \frac{
    \sum\limits_{(i,q) \in A} w_q w_{i} \, \delta_{i}
    }{
	\sum\limits_{(i,q) \in A} w_q w_{i}
    }.
    \label{equation:cf_xcf}
\end{equation}
Here, $i$ and $j$ represent pixels on the wavelength grid of the Ly$\alpha$ forest . The weights $w_i,w_q$ are defined as
\begin{equation}
\begin{aligned}
    w_{i}(\lambda) =\left( \frac{1 + z_\lambda^{i}}{1 + z_0} \right)^{\gamma_\alpha -1}\times\frac{1}{\left(\sigma_{\delta}(\lambda) \right)^2}
,
\end{aligned}\label{equa:Lya_weight}
\end{equation}
and 
\begin{equation}
        w_q = \left( \frac{1 + z_q}{1 + 2.25} \right)^{\gamma_q -1}.\label{equa:QSO_weight}
\end{equation}
Here $\gamma_\alpha=2.9$ \cite{mcdonald2006lyalpha} and $\gamma_q=1.44$ \cite{des2020completed} are two parameters related to redshift evolution of the Ly$\alpha$ and quasar biases, respectively. Although these two parameter values were originally optimized for $z>2$, we retain them in our analysis because tests with alternative choices showed no significant impact on the resulting correlation functions. $z_\lambda^{i}$ refers to the redshift of a given Ly$\alpha$ forest absorption at the $i_{\rm{th}}$ wavelength pixel, $z_q$ is the quasar redshift, and $\sigma_{\delta}$ account for the overall noise, defined in Eq.~\ref{equa:sigma_delta}. For ELG-forest cross correlation, we use the same weight definition as Eq.~\ref{equa:QSO_weight}.

The covariance matrix of the 
auto- and cross-correlations are calculated by sub-sampling, in the same way as described in Sections 3.2 and 3.3 of \cite{des2020completed}.
The method consists of measuring the correlation
functions in different regions of the sky
and deducing the covariances from the variations
between different regions.

\subsection{Modeling and fitting of BAO}
\label{sec:model_cf}

We fit the measured Ly$\alpha$ auto- and cross-correlation functions
following a similar model to those used in recent eBOSS and DESI 
Ly$\alpha$ analyses \cite{des2020completed,desi2024lya,abdul2025desi,gordon2023desi}.
In Fourier space, the model power spectra for tracers 
$i,j\in\{\mathrm{Ly}\alpha,\mathrm{QSO},\mathrm{ELG}\}$ are written as
\begin{equation}
\begin{aligned}
    P_{ij}(k,\mu)
    = &\,b_ib_j\,
      \bigl(1+\beta_i\mu^2\bigr)\,
      \bigl(1+\beta_j\mu^2\bigr)\\
      &\times P_L(k)\,
      F_{\rm NL,ij}(k,\mu)\,
      G(k_\parallel,k_\perp),
\end{aligned}
\label{eq:Pij_model}
\end{equation}
where $P_L(k)$ represents the linear matter power spectrum, $b_i$ and $\beta_i$ are the linear bias and Kaiser redshift-space distortion (RSD) parameter of tracer $i$, and $k^2=k_\parallel^2+k_\perp^2$ and $\mu=k_\parallel/k$.
$F_{\rm NL,ij}(k,\mu)$ collects the small-scale corrections calibrated for Ly$\alpha$ forest
clustering, which we keep fixed in this work according to
the default prescription adopted in \cite{gordon2023desi}. For the cross-correlations, we do not include any additional tracer
peculiar velocity or redshift-error. Following the DESI Ly$\alpha$ BAO analysis \cite{desi2024lya}, the model is
evaluated in each $(r_\perp,r_\parallel)$ bin at the corresponding weighted
mean redshift (computed with the same weights defined in Eq.~\ref{equa:Lya_weight} and Eq.~\ref{equa:QSO_weight}).
All bias parameters are defined at an effective redshift,
$z_{\rm eff}=1.59$ for the joint ($\mathrm{Ly}\alpha\times\mathrm{Ly}\alpha$,
$\mathrm{Ly}\alpha\times\mathrm{QSO}$) fit and $z_{\rm eff}=1.58$ for the joint
($\mathrm{Ly}\alpha\times\mathrm{Ly}\alpha$, $\mathrm{Ly}\alpha\times\mathrm{ELG}$) fit.
The redshift evolution of bias amplitudes follows the same power-law prescription
as in \cite{desi2024lya}.
We assume $\beta_{\mathrm{Ly}\alpha}$ to be redshift-independent
and set $\beta_g=f/b_g$ for the discrete tracers.
The function $G(k_\parallel,k_\perp)$ accounts for the finite bin size used to 
measure the correlation functions. Following \cite{gordon2023desi}, it is 
implemented as
\begin{equation}
    G(k_\parallel,k_\perp)
    = \mathrm{sinc}\!\left(\frac{k_\parallel\Delta r_\parallel}{2}\right)
      \mathrm{sinc}\!\left(\frac{k_\perp\Delta r_\perp}{2}\right),
\end{equation}
In addition to this explicit bin-averaging, the finite resolution of the 
\textsc{CoLoRe} density field ($d x\simeq 7.9\,h^{-1}{\rm Mpc}$) introduces additional 
small-scale smoothing,
which we capture through two free parameters $(\sigma_{\parallel}^{\rm smooth},
\sigma_{\perp}^{\rm smooth})$ that multiply Eq.~\eqref{eq:Pij_model} via
\begin{equation}
\begin{aligned}
D_{\rm smooth}(k,\mu)
&= \exp\biggl[
    -\tfrac12 k^2 \bigl(
        \mu^2 \sigma_{\parallel,\rm smooth}^{2}
\\
&\qquad\qquad
        +(1-\mu^2)\sigma_{\perp,\rm smooth}^{2}
    \bigr)
\biggr].
\end{aligned}
\end{equation}

The configuration-space correlation functions used in the BAO fits are 
obtained by Fourier transforming Eq.~\eqref{eq:Pij_model}. 
To isolate the BAO signal, the correlation function is artificially split into two parts
\cite{kirkby2013fitting}
\begin{equation}
\begin{aligned}
\xi(r_\parallel,r_\perp,\alpha_\parallel,\alpha_\perp)
     = &\,\,\xi_{\rm sm}(r_\parallel,r_\perp)
    \\&+ A_{\rm BAO}\,
      \xi_{\rm peak}(\alpha_\parallel r_\parallel,\alpha_\perp r_\perp),
\end{aligned}
\end{equation}
where $\xi_{\rm sm}$ and $\xi_{\rm peak}$ are Fourier transforms of the 
smooth and oscillatory components of the model power spectrum. For $\xi_{\rm peak}$, we add three free parameters.
The dilation parameters
\begin{equation}
    \alpha_\parallel =
    \frac{[D_H(z_{\rm eff})/r_d]}
         {[D_H(z_{\rm eff})/r_d]_{\rm fid}},
    \quad
    \alpha_\perp =
    \frac{[D_M(z_{\rm eff})/r_d]}
         {[D_M(z_{\rm eff})/r_d]_{\rm fid}}
\end{equation}
encode the anisotropic rescaling relative to the above-mentioned fiducial cosmology, for the BAO peak.  
The parameter $A_{\rm BAO}$ regulates the overall amplitude of the BAO peak.

Parameter estimation and fitting of the correlation functions are performed with the \textsc{vega} framework,\footnote{\url{https://github.com/andreicuceu/vega}} \cite{cuceu2023alcock}
using the covariance matrices we computed for the noisy CSST-like mocks
(Sec.~\ref{sec:measure_cf+fitting}).  
The free parameters in our fits are therefore 
$(\alpha_\parallel,\alpha_\perp,A_{\rm BAO})$, 
$(b_{\rm Ly\alpha},\beta_{\rm Ly\alpha})$, 
$(\sigma_{\parallel}^{\rm smooth},\sigma_{\perp}^{\rm smooth})$, 
and the tracer biases $b_{\rm QSO}$ or $b_{\rm ELG}$ for the cross-correlations.

\section{Results}\label{sec:results}

\begin{figure*}
\centering
\includegraphics[width=\textwidth]{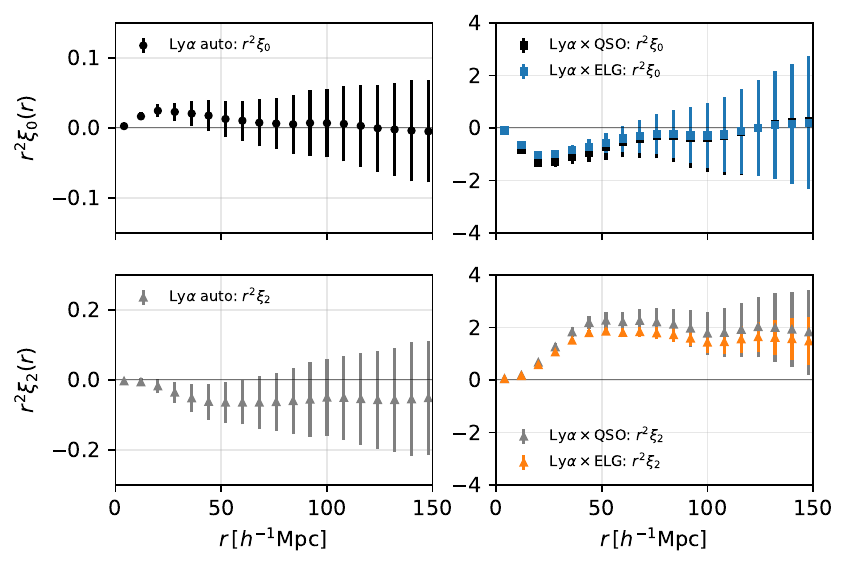}
\caption{Multipoles of the CSST Ly$\alpha$ forest correlation functions used in our forecast.
The points show the fiducial (noiseless) correlation functions adopted as the ``true'' data vector
in the likelihood, compressed into the monopole ($\ell=0$) and quadrupole ($\ell=2$) as $r^2\xi_\ell(r)$.
The error bars indicate the $1\sigma$ uncertainties derived from the covariance matrix estimated
from the noisy CSST-like mocks described in Sec.~\ref{sec:measure_cf+fitting}.
Left panel: Ly$\alpha\times$Ly$\alpha$ auto-correlation at $z_{\rm eff}=1.59$.
Right panel: Ly$\alpha\times$QSO cross-correlation at $z_{\rm eff}=1.59$ and Ly$\alpha\times$ELG cross-correlation at $z_{\rm eff}=1.58$.}
\label{fig:cf}
\end{figure*}

\begin{table*}[htbp]
  \centering
  \caption{Forecasted constraints on BAO, Ly$\alpha$ and tracer parameters
  from the MCMC. For each parameter we quote the posterior mean and
  68\% credible interval. When $A_{\rm BAO}$ or
  $(\alpha_{\parallel},\alpha_{\perp})$ are not fitted, they are
  fixed to unity. The effective redshift is
  $z_{\rm eff}=1.59$ for the Ly$\alpha$–QSO configurations and
  $z_{\rm eff}=1.58$ for the Ly$\alpha$–ELG configurations. Note that $b_g=b_{\rm{QSO}}$ for quasars and $b_g=b_{\rm{ELG}}$ for ELGs. In the specific case of fitting $(\alpha_{\parallel},\alpha_{\perp})$ with Ly$\alpha\times$QSO, a prior with a width of 0.4 was added to these parameters.}
  \label{tab:big_mcmc_summary}
  \resizebox{0.97\textwidth}{!}{
  \begin{tabular}{lcccc}
\hline\hline
Parameter & Ly$\alpha\times$Ly$\alpha$ + Ly$\alpha\times$QSO & Ly$\alpha\times$Ly$\alpha$ + Ly$\alpha\times$QSO  & Ly$\alpha\times$Ly$\alpha$ + Ly$\alpha\times$ELG & Ly$\alpha\times$Ly$\alpha$ + Ly$\alpha\times$ELG  \\
\hline
$A_{\mathrm{BAO}}$ & $0.96^{+0.39}_{-0.39}$ & $\cdots$ & $1.00^{+0.27}_{-0.27}$ & $\cdots$ \\
$\alpha_{\mathrm{iso}}$ & $\cdots$ & $1.02^{+0.11}_{-0.10}$ & $\cdots$ & $1.01^{+0.07}_{-0.06}$ \\
$b_{\mathrm{Ly}\alpha}$ & $-0.029^{+0.007}_{-0.007}$ & $-0.029^{+0.007}_{-0.007}$ & $-0.029^{+0.006}_{-0.006}$ & $-0.028^{+0.006}_{-0.006}$ \\
$\beta_{\mathrm{Ly}\alpha}$ & $2.00^{+0.60}_{-0.47}$ & $1.98^{+0.57}_{-0.45}$ & $2.05^{+0.35}_{-0.32}$ & $2.06^{+0.36}_{-0.30}$ \\
$\sigma_{\perp}^{\mathrm{smooth}}$ & $6.92^{+0.75}_{-0.89}$ & $7.38^{+0.69}_{-0.77}$ & $6.99^{+0.35}_{-0.36}$ & $7.01^{+0.34}_{-0.36}$ \\
$\sigma_{\parallel}^{\mathrm{smooth}}$ & $6.74^{+1.69}_{-2.29}$ & $7.23^{+1.61}_{-2.05}$ & $7.03^{+0.85}_{-0.93}$ & $7.06^{+0.85}_{-0.89}$ \\
$b_g$ & $2.57^{+0.57}_{-0.40}$ & $2.64^{+0.56}_{-0.39}$ & $1.94^{+0.43}_{-0.29}$ & $1.95^{+0.44}_{-0.30}$ \\
\hline\hline
\end{tabular}

  }
\end{table*}

The purpose of this study is to forecast the measurement of Ly$\alpha$ correlation functions using CSST mock data and the ability of constraining Ly$\alpha$ biases and the BAO signal. To quantify the expected sensitivity, we use the following approach: using the fiducial set of Ly$\alpha$ and tracer parameters, as well as fiducial cosmology, we compute a model two-dimensional correlation function as described in Sec.~\ref{sec:model_cf}; this model is taken as the true data vector. On the other hand, we use the covariance matrix computed from the noisy mock, as described in Sec.~\ref{sec:measure_cf+fitting}. From this "noiseless" data vector and covariance, we build a Gaussian likelihood and sample the posterior
distribution of the model parameters. This is done using the \textsc{vega}
software, which relies on 
\textsc{PolyChord}\footnote{\url{https://github.com/PolyChord}} as a
nested sampler. From the resulting MCMC chains, we can forecast statistical precision and parameter degeneracies expected for the CSST
survey.

Figure~\ref{fig:cf} presents multipoles of the Ly$\alpha$
auto- and cross-correlations for the "noiseless" data vector (shown as the means), with errorbars computed using covariance matrix from the noisy mock. The monopole and quadrupole are 
obtained by projecting the measured two-dimensional correlation
functions $\xi(r_\parallel,r_\perp)$ onto Legendre multipoles and
plotted as $r^2\xi_\ell(r)$. These compressed statistics are used only for visualization:
all parameter constraints reported below are derived from fits to the
full two-dimensional correlation functions $\xi(r_\parallel,r_\perp)$.

We fit jointly the
Ly$\alpha$ auto-correlation and the Ly$\alpha$–tracer cross-correlation
over $25<r<180\,h^{-1}{\rm Mpc}$ with four configurations. On the one hand, we consider either an ``amplitude'' fit in which
$A_{\rm BAO}$ is allowed to vary while
$(\alpha_{\parallel},\alpha_{\perp})$ are fixed to unity, or an anisotropic BAO fit in which
$(\alpha_{\parallel},\alpha_{\perp})$ are free parameters while
$A_{\rm BAO}$ is fixed to unity. On the other hand, we consider either QSO or ELG as tracers for the Ly$\alpha$-tracer cross-correlation. The forecasted posterior constraints are
summarized in Table~\ref{tab:big_mcmc_summary}. For each parameter we
report the posterior median and 68\% credible interval inferred from the
MCMC chains.

Using Ly$\alpha$ auto- and Ly$\alpha$–QSO
cross-correlations at $z_{\rm eff}=1.59$, the BAO amplitude fit yields 
$A_{\rm BAO}=0.96^{+0.39}_{-0.39}$: we forecast a marginal
$2.5\sigma$ detection of the BAO peak. In the corresponding fit with fixed $A_{\rm BAO}$ and free dilation parameters, given the weakness of the BAO signal we have to add a wide prior on both $\alpha_\parallel$ and $\alpha_\perp$, centered at 1.0 and with a width of 0.4. In that case we recover a $10\%$ precision on  $\alpha_\parallel$ and $17\%$ precision on $\alpha_\perp$. We derive the isotropic scaling parameter and obtain $\alpha_{\rm iso} = \left(\alpha_\perp^2 \alpha_\parallel\right)^{1/3}= 1.01^{+0.10}_{-0.09}$. Replacing QSOs by ELGs in the
cross-correlation gives a higher BAO
significance of $3.7\sigma$, such that no prior on $\alpha_\parallel$ and $\alpha_\perp$ is needed. We obtain a tighter constraint on the BAO scale $\alpha_{\rm iso} = 1.01^{+0.07}_{-0.06}$.

Considering other fit parameters, we first notice that the smoothing parameters are constrained around
$\sigma_{\perp}^{\rm smooth}\simeq
\sigma_{\parallel}^{\rm smooth}\simeq 7\,h^{-1}{\rm Mpc}$, consistent
with the effective resolution expected from the
\textsc{CoLoRe} grid spacing.
Depending on the fit configuration, the forecasted precision for the Ly$\alpha$ bias parameter $b_{\mathrm{Ly}\alpha}$ is
20–30\%. The precision for the Ly$\alpha$ RSD parameter $\beta_\alpha$ is similar, ranging from 15 to 30\%. This is while fitting also the tracer biases, whose best-fit values are consistent with the input model. On the other hand, we have also explored fits in which external priors are imposed on
$b_{\mathrm{QSO}}$ and $b_{\mathrm{ELG}}$ (up to fixing them to their
input values). In this case, we found improved constraints on $b_{\mathrm{Ly}\alpha}$ by about a factor of two, to a
fractional precision of $10\%$. This is consistent with the fact that, in terms of signal-to-noise, the cross-correlation measurements dominate the auto-correlation, as can be seen in Fig.~\ref{fig:cf}.

Together, these
forecasts indicate that CSST can deliver the first measurements of the
large-scale Ly$\alpha$ bias and redshift-space anisotropy below $z=2$ and provide complementary BAO measurements for $1.5<z<2$.

\section{Discussion and conclusion}\label{sec:discussion_conclusion}
Our forecasts demonstrate that CSST, despite its modest spectral resolution ($R=241$) and limited exposure time in the wide survey ($600$\,s), can provide marginal BAO constraints and measure the Ly$\alpha$ forest bias in the range $1<z<2$. This will complement existing measurements at higher redshift from eBOSS and DESI, filling an important observational gap in the study of the IGM.

We forecast a precision of 10\% (7\%) on the isotropic BAO scale from Ly$\alpha\times$Ly$\alpha$ combined with Ly$\alpha\times$QSO (ELG), corresponding to a $ 2.5\sigma$ ($3.7\sigma$) detection of the BAO peak at $z_{\rm eff}=1.59$ (1.58). The Ly$\alpha$ bias parameters can also be measured with 10 to 30~\% precision, depending on priors on other tracer's biases. This establishes CSST as a useful probe of low-redshift Ly$\alpha$ forest cosmology. 

Because the SNR of the Ly$\alpha$ forest in
our CSST configuration is modest, the Ly$\alpha$ auto-correlation mainly
constrains the overall clustering amplitude and redshift-space
distortions of the forest. In contrast, the Ly$\alpha$–QSO and
Ly$\alpha$–ELG cross-correlations benefit from the higher bias of the
tracer populations and therefore carry most of the BAO information.

Several systematic effects require further investigation. First, we did not take into account any astrophysical contaminants in our analysis, such as Damped Ly$\alpha$ Absorbers (DLAs), metal lines etc, that are known to have non-negligible impact on Ly$\alpha$ analysis. 
Second, we anticipate that quasar redshift errors and mis-targeting, as well as continuum fitting uncertainties will impact our present treatment.

On the other hand, we limited our analysis to $z<2$ in order to highlight the complementarity with ground-based surveys, but CSST will also observe a substantial number of quasars at $z>2$ with higher SNR, allowing BAO measurements at $z_{\rm eff}\sim1.8$ with potentially better precision than reported here.
Looking forward, CSST slitless spectroscopy will also enable the construction of a large catalog of high column density H\,\textsc{i} systems, including damped Ly$\alpha$ absorbers. Auto-correlations of DLAs and their cross-correlations with Ly$\alpha$ transmission, quasars, and ELGs have the potential to provide an independent BAO measurement and additional consistency checks on systematics affecting the Ly$\alpha$ forest. 

In summary, this work presents a first end-to-end forecast of three-dimensional Ly$\alpha$ forest clustering with CSST at $z<2$. Within a CSST-like survey configuration, we find that Ly$\alpha$ auto- and Ly$\alpha$–tracer cross-correlations can be measured with sufficient signal-to-noise to constrain the large-scale Ly$\alpha$ bias and redshift-space distortions, and to extract BAO information at the $\sim10\%$ level. These results demonstrate the feasibility of low-redshift Ly$\alpha$ forest analyses with CSST and motivate more detailed studies, including more realistic mock catalogs and a systematic treatment of observational and astrophysical effects, in preparation for the survey data.

\section*{Acknowledgements}
We thank Andreu Font-Ribera for discussing the extrapolation of \textsc{LyaCoLoRe}, as well as Andrei Cuceu on the \textsc{Vega} software. We thank James Rich for providing useful suggestions. We acknowledge support from the ANR grant ANR-22-CE92-0037.

\end{multicols}
\end{document}